\documentstyle[twocolumn,aps,prc,epsf]{revtex}
\begin{document}

\twocolumn[\hsize\textwidth\columnwidth\hsize\csname
@twocolumnfalse\endcsname

\draft
\title{Search for excited states in $\bbox{^3}$H and
$\bbox{^3}$He}
\author{Attila Cs\'ot\'o$^{1,2}$ and G.~M. Hale$^1$}
\address{$^1$Theoretical Division, Los Alamos National
Laboratory, Los Alamos, New Mexico 87545\\
$^2$Department of Atomic Physics, E\"otv\"os University,
Puskin u.\ 5-7, H--1088 Budapest, Hungary}
\date{August 26, 1998}

\maketitle

\begin{abstract}
\noindent
The $d+N$ systems are studied in a three-body model, using
phenomenological N--N interactions. The scattering matrices
are calculated by using the Kohn-Hulth\'en variational method.
Then, they are analytically continued to complex energies and
their singularities are localized. We find a virtual state at
$E=-1.66$ MeV in $^3$H and a pair of states at $E=(-0.42\pm i0.52)$
MeV in $^3$He relative to the $d+N$ thresholds, respectively. In 
addition, we discuss some general aspects and problems of virtual 
states which may be useful also in the study of other systems, 
like $^{10}$Li. 
\end{abstract}
\ \\
]

\narrowtext

\section{Introduction}

The lightest nucleus with a well-established spectrum of
excited states is $^4$He. There have been numerous attempts,
both experimental and theoretical, to identify excited
states in $A=3$ nuclei; for an extensive review see Ref.\
\cite{Moller}. Although there are indications for
resonance-like behavior in the $3n$ and $3p$ systems in
certain experiments \cite{3ny}, others do not seem to
support these findings \cite{3nn}. Theoretically the most
well-established case is a broad three-neutron (and
three-proton) resonance in the $J^\pi=3/2^+$ partial wave
\cite{3n}. However, the existence of these structures has
not been confirmed by experiments yet.

According to Ref.\ \cite{Moller}, no unambiguous
experimental evidence has been found to support the
existence of low-lying $^3$H and $^3$He resonances. However,
several calculations, mostly confined to small model spaces,
indicate a virtual state in $^3$H, and corresponding
subthreshold resonances in $^3$He. Recently, the authors of
a ${\rm H}({^6{\rm He}},{^4{\rm He}})$ experiment have
claimed to find an excited state of $^3$H at $E^*=7\pm0.3$
MeV excitation energy with $\Gamma=0.6\pm0.3$ MeV width
\cite{Aleksandrov}. According to one interpretation
\cite{Barabanov} this is a $1/2^+$ excited state formed
mainly in the $(nn)p$ channel. In the mirror $^3$He nucleus
the Ref.\ \cite{Aleksandrov} experiment does not indicate an
excited state. However, in a recent work \cite{Kievsky} it
was shown that the effective range function of the doublet
$d+p$ scattering has a singularity at a negative energy,
just below the $d+p$ threshold. This would indicate the
existence of subthreshold virtual states in $^3$He.

In the present work we study the $^3$H and $^3$He nuclei
with the aim to clarify some of the possibilities mentioned
above. We describe the $d+N$ scattering in a nonrelativistic
three-body potential model and study the analytic properties
of the scattering matrices. Since virtual states play an
important role in our study, we briefly discuss some of their
features in Sec.\ II, and show some problems associated with
their experimental investigation. Our three-body calculations
are reviewed in Sec.\ III, while Sec.\ IV offers some conclusions.

\section{Virtual states in scattering theory}

The spectrum of a quantum scattering system generally
consists of bound states and continuum states. The continuum
has structure caused by the complex-energy singularities of
the scattering matrix. For a broad class of physically
interesting interactions all singularities within the
meromorphic region of the potential are physical and
correspond to solutions of the Schr\"odinger equation with
purely outgoing asymptotics. The following types of
singularities are usually defined on the complex $k$ (wave
number) plane: i) bound states with $k_B=i\gamma$
($\gamma>0$); ii) virtual (antibound) states with
$k_A=-i\gamma$ ($\gamma>0$); resonances with
$k_D=\pm\kappa-i\gamma$ ($\kappa,\,\gamma>0;\ \kappa>\gamma$);
and quasi-resonances with $k_C=\pm\kappa-i\gamma$
($\kappa,\,\gamma>0;\ \kappa<\gamma$).

Our main interest in the present work is in the virtual
states. Interestingly, although a virtual state is an
unbound state with exponentially increasing wave function,
it corresponds to negative energy,
$E_V=k^2\hbar^2/2m=-\gamma^2\hbar^2/2m$. The cross section
of a process which involves a virtual state behaves like
$\sigma(E)\sim1/(E+\vert E_V\vert)$ for $E>0$. Thus it is
singular at the unphysical negative energy $E=-\vert
E_V\vert$, and increases with decreasing positive energies.
This behavior can have rather strange effects on
experimental measurements. If a measurement is performed on
a system which has a virtual state and the interaction-free
cross section (phase space) of the measured process drops to
zero with decreasing energy, then the result of the
measurement is a cross section with a low-energy peak. This
may be interpreted as a low-energy resonance at positive
energies. However, the peak cannot be fitted with a
Breit-Wigner form, which causes the extracted ``resonance
parameters'' to be highly dependent on the fit procedure. A
good example of this experimental difficulty concerning
virtual states is $^{10}$Li. We believe that the experimental
confusion regarding the nature of the ground state of
$^{10}$Li \cite{Benenson} is caused by the fact that there
is a virtual state in the $^9{\rm Li}+n$ system
\cite{Thompson}. The energy of this state is negative, for instance
$E_V\approx -0.03$ MeV for the P2 interaction of Ref.\
\cite{Thompson}. If, however, the experimental procedures
assume a low-energy resonance in this system and try
to fit the various measured cross sections accordingly by
Breit-Wigner forms, then they can get rather different
results.

\begin{figure}
\epsfxsize 8cm \epsfbox{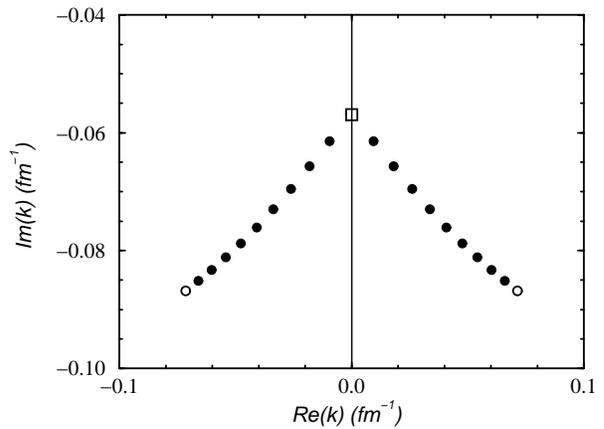}
\caption{Trajectories of the $^1S_0$ $N+N$ $S$-matrix poles.
The open square corresponds to the $n+n$ and $n+p$ poles,
wile the open circles denote the pair of conjugate poles in
the $p+p$ system. The filled circles come from calculations
where $c\cdot V_{\rm Coul}^{pp}$ is added to the $n+n$
interaction ($0<c<1$).}
\label{traj}
\end{figure}

We would like to emphasize that pure virtual states with
pure imaginary wave number can only exist in neutral
systems with two-body asymptotics and only in $S$-wave states. By making the
interaction more attractive the virtual state can be moved
from the negative imaginary $k$-axis to the positive one,
ending up with a bound state \cite{Glockle}. If, however, a
Coulomb or centrifugal barrier is added to the interaction,
then the virtual pole moves from the imaginary axis into the
complex $k$ plane resulting in a quasi-resonance. In
order to illustrate this process we show in Fig.\ \ref{traj}
the pole trajectory corresponding to the $^1S_0$ $N+N$
system. We start from $n+n$ and smoothly switch on the $p+p$
Coulomb interaction to finally end up as the $p+p$ system.
One can see that the effect of the Coulomb force is to move
the virtual pole into the complex plane (and to create a
conjugate pole) as discussed above. We use here the
Eikemeier-Hackenbroich (EH) N--N interaction \cite{EH}. The
$n+n$ and $p+p$ poles are at $E_{nn}=-0.134$ MeV and
$E_{pp}=(-0.101\pm i0.515)$ MeV energies respectively. These
numbers are in reasonable agreement with those that can be
extracted from experimental data, $E_{nn}=-0.123$ MeV and
$E_{pp}=(-0.140\pm i0.467)$ MeV \cite{Kok}. Our interaction is
charge-independent, so it gives the same pole position for
$n+n$ and $n+p$. Naturally, it cannot reproduce the
experimental $E_{np}=-0.066$ MeV \cite{Kok}.

We note that although one cannot have pure virtual
poles in charged systems, a pair of quasi-resonance poles close to
the negative imaginary $k$ axis can have observable effects
at positive real energies, like in the case of $p+p$. The
reason for this is the fact that unlike in the case of
usual resonance poles, both of such poles are roughly the
same distance from the real energy axis of the physical
energy sheet. This is not the case for usual resonant states. The
$\varepsilon=E_r-i\Gamma/2$ ($E_r,\Gamma>0$) main pole of a
usual resonance can be reached by simply crossing the real
energy axis and going to the fourth quadrant of the energy
plane. However, the conjugate pole can only be reached
through a long path by crossing the real energy axis, going
into the fourth quadrant, and then encircling the origin and
ending up at $\varepsilon=E_r+i\Gamma/2$ ($E_r,\Gamma>0$).

To recap our brief discussion of virtual states, we
emphasize the sometimes forgotten fact that the energy
corresponding to these states is negative. We also note that
in charged systems pure virtual states cannot exist. However,
the effect of a pair of conjugate quasi-resonances close
to the negative imaginary $k$ axis (corresponding to a
virtual state in a neutral system) can have significant
observable effects.

\section{Three-body calculations for \lowercase{d}+N
scattering}

We solve the three-body Schr\"odinger equation for $^3$H and
$^3$He using the EH N--N interaction. This interaction
gives a good general description of the $N+N$ scattering
data and the deuteron properties \cite{EH}. Technical
problems prevent us from using the most modern interactions,
but for our current purposes the EH interaction is quite
satisfactory. In our calculations the same nucleon mass
$M_N=(M_p+M_n)/2$ is used for both protons and neutrons.

In this paper we study only $J^\pi=1/2^+$ states. As a first
step we solve the bound-state problem of $^3$H and $^3$He
using the variational Gaussian-basis coupled rearrangement
channel method \cite{Kamimuratrit}. We include all 23
channels with $l_1,l_2\leq2$ in the $\bigg [\Big
[(S_1,S_2)S_{12},S_3\Big]S,(l_1,l_2)L\bigg ]J^\pi$ coupling
scheme. Here $S_1=S_2=S_3=1/2$ are the nucleon spins,
$S_{12}$ is the coupled two-nucleon intrinsic spin, $S$ is
the the total intrinsic spin, $l_1$ and $l_2$ are the
orbital angular momenta of the two relative motions,
respectively, $L$ is the total orbital angular momentum, $J$
is the total spin, and $\pi=(-1)^{l_1+l_2}$ is the parity.
Our calculated binding energies, $E_{^3{\rm H}}=-7.65$ MeV and $E_{^3{\rm
He}}=-6.99$ MeV, are smaller than the experimental values
$E_{^3{\rm H}}^{\rm Exp}=-8.482$ MeV and $E_{^3{\rm He}}^{\rm Exp}=
-7.718$ MeV, respectively. Our triton binding energy is close
to those coming from calculations that use the most modern
two-nucleon forces \cite{Nogga}. Thus, a large part of the underbinding
can be attributed to the lack of three-body forces in our
model.

We find that by keeping only the 9 most important channels, shown
in Table \ref{space1}, the binding energies change by only
0.05 MeV. In the following we keep only these 9 channels.
In order to have a coupled ${^3S}_1-{^3D}_1$ deuteron in our
asymptotic $d+N$-type wave function components, we use the
$\Bigg [\bigg [\Big [(S_1,S_2)S_{12},l_1\Big ]I_1,S_3\bigg
]I,l_2\Bigg ]J^\pi$ coupling scheme in the scattering
calculations. Here $I_1$ is the total (intrinsic plus
orbital) spin of the two-nucleon subsystem and $I$ comes
from the coupling of $I_1$ and $S_3$. We show our model space
in this coupling in Table \ref{space2}. Hence, we have 7
channels: two channels containing a ${^3S}_1-{^3D}_1$
deuteron plus a nucleon with $l_2=0$ and 2, respectively
(lines 1--2 and 3--4 in Table \ref{space2}), three channels
with $^3D$ states inside the $n+p$ system (lines 5,6,
and 7 in Table \ref{space2}), and two channels with $^1S_0$
$n+p$ and $n+n$ ($p+p$ for $^3$He) two-body subsystems, respectively
(lines 8 and 9 in Table \ref{space2}). The multichannel
scattering problem is solved by using the Kohn-Hulth\'en
variational method for the $S$ matrix \cite{Kamimura}. In
all our calculations we remain below the three-body breakup
threshold. It means that in channels 5--9 in Table
\ref{space2} the wave functions have three-body bound-state
asymptotics \cite{Merkurev}. In these channels we keep the
same Gaussian basis as in the bound state calculations.

The scattering matrices, coming from the Kohn-Hulth\'en
calculations, are analytically continued to the multisheeted
complex Riemann energy surface using the methods of Ref.\
\cite{cont}, and their singularities are localized. As a
test calculation we searched for the bound-state poles and
found them at the same energies as in the bound-state
calculations. In addition, a virtual state is found in $^3$H at
$E^{^3{\rm H}}_V= -1.66$ MeV, relative to the $d+n$ threshold. We note
that as our triton ground state is underbound so is probably
this virtual state. The effect of underbinding in a virtual
state is that $\vert E_V\vert$ becomes {\it larger}
\cite{Glockle}. This means that if our $N-N$ interaction reproduced
the correct binding energy of $^3$H and $^3$He then the
virtual-state pole would be closer to zero energy. In $^3$He we find a
pair of conjugate poles at $E^{^3{\rm He}}_V=(-0.42\pm i0.52)$ MeV,
relative to the $d+p$ threshold. The insufficient attraction
in our model probably moves the poles too far away from the
negative energy axis.

A practical way to extract the parameters of the virtual states of
$^3$H and $^3$He from experimental data is to describe these nuclei
within the extended $R$-matrix model \cite{Hale1}, which can be
continued to complex energies. An extensive analysis of the $A=3$ data
is being performed using this method \cite{Hale2}. Preliminary results
indicate the presence of $^2S_{1/2}$ $^3$H and $^3$He virtual states
at $E^{^3{\rm H}}_V= -1.07$ MeV and $E^{^3{\rm He}}_V=(-0.72\pm i0.23)$
MeV, respectively. The results of our three-body calculations are
consistent with these findings if we take into account the fact that
the insufficient binding probably increases $\vert E^{^3{\rm
H}}_V\vert $ and pushes the $^3$He pole away from the negative axis.

In comparing our results to previous calculations, we note that in
Ref.\ \cite{Friar}, negative-energy poles were found in the
$^2S_{1/2}~k {\rm cot}\delta$ function for $d+n$ at $\approx -160$
keV and for $d+p$ at $\approx -25$ keV, using the $S$-wave MT I-III
interaction of Ref.\ \cite{MT}. A similar pole was found in the $d+p$
phase-shift analysis of Black et al.\ \cite{Black}.  These
real-energy poles in the $d+N$ effective-range expansion are
clearly related to the presence of the virtual states we have
found, as will be discussed in a later publication \cite{Hale2}.  

\section{Conclusion}

In summary, we have localized virtual $d+N$ states in $^3$H and
$^3$He in a model that contains all relevant three-body channels.
These states can be considered as special excited states of the $A=3$
nuclei. Our results undoubtedly show the existence of these
structures, although their properties could change somewhat
if the most modern N--N interactions with three-body forces were
used.  Such calculations are, however, beyond the scope of the
present work.

\acknowledgments

This work was performed under the auspices of the U.S.\
Department of Energy. Some of the calculations were
performed at the Physics Department of Aarhus University,
Denmark. A.\ C.\ thanks K.\ Langanke for his help and
the Danish Research Council and the Theoretical
Astrophysics Center for partial financial support. This
work was supported by OTKA Grant.\ F019701 and by the 
Bolyai Fellowship of the Hungarian Academy of Sciences.

\narrowtext
\begin{table}
\caption{Channel configurations used in our 9-channel
calculations in the $\bigg [\Big
[(S_1,S_2)S_{12},S_3\Big]S,(l_1,l_2)L\bigg ]J^\pi$ coupling
scheme.}
\begin{tabular}{cccccc}
Configuration & $S_{12}$ & $S$ & $l_1$ & $l_2$ & $L$  \\
\tableline
$N(pn)$ & 1 & 1/2 & 0 & 0 & 0 \\
$N(pn)$ & 1 & 3/2 & 2 & 0 & 2 \\
$N(pn)$ & 1 & 3/2 & 0 & 2 & 2 \\
$N(pn)$ & 1 & 1/2 & 2 & 2 & 0 \\
$N(pn)$ & 1 & 1/2 & 2 & 2 & 1 \\
$N(pn)$ & 1 & 3/2 & 2 & 2 & 1 \\
$N(pn)$ & 1 & 3/2 & 2 & 2 & 2 \\
$N(pn)$ & 0 & 1/2 & 0 & 0 & 0 \\
$p(nn)$ or $n(pp)$ & 0 & 1/2 & 0 & 0 & 0 \\
\end{tabular}
\label{space1}
\end{table}

\narrowtext
\begin{table}
\caption{Channel configurations used in our 9-channel
calculations in the $\Bigg [\bigg [\Big [(S_1,S_2)S_{12},
l_1\Big ]I_1,S_3\bigg ]I,l_2\Bigg ]J^\pi$ coupling scheme.}
\begin{tabular}{cccccc}
Configuration & $S_{12}$ & $l_1$ & $I_1$ & $I$ & $l_2$ \\
\tableline
$N(pn)$ & 1 & 0 & 1 & 1/2 & 0 \\
$N(pn)$ & 1 & 2 & 1 & 1/2 & 0 \\
$N(pn)$ & 1 & 0 & 1 & 3/2 & 2 \\
$N(pn)$ & 1 & 2 & 1 & 3/2 & 2 \\
$N(pn)$ & 1 & 2 & 2 & 3/2 & 2 \\
$N(pn)$ & 1 & 2 & 2 & 5/2 & 2 \\
$N(pn)$ & 1 & 2 & 3 & 5/2 & 2 \\
$N(pn)$ & 0 & 0 & 0 & 1/2 & 0 \\
$p(nn)$ or $n(pp)$ & 0 & 0 & 0 & 1/2 & 0 \\
\end{tabular}
\label{space2}
\end{table}

\narrowtext


\begin{references}
\bibitem{Moller} K. M\"oller and Yu.~V. Orlov, Sov. J. Part.
Nucl. {\bf 20}, 569 (1989).
\bibitem{3ny} J. Sperinde, D. Frederickson, R. Hinkins,
V. Perez-Mendez, and B. Smith, Phys. Lett. {\bf 32B},
185 (1970); L.~E. Williams, C.~J. Batty, B.~E. Bonner,
C.~Tschal\"ar, H.~C. Ben\"ohr, and A.~S. Clough, Phys. Rev.
Lett. {\bf 23}, 1181 (1969); A. Stetz {\it et al.}, Nucl.
Phys. {\bf A457}, 669 (1986).
\bibitem{3nn} M. Yuly {\it et al.}, Phys. Rev. C {\bf 55},
1848 (1997); M. Palarczyk {\it et al.}, Phys. Rev. C {\bf 58}, 
645 (1998).
\bibitem{3n} A. Cs\'ot\'o, H. Oberhummer, and R. Pichler,
Phys. Rev. C {\bf 53}, 1589 (1996).
\bibitem{Aleksandrov} D.~V. Aleksandrov, E.~Yu. Nikol'skii,
B.~G. Novatskii, and D.~N. Stepanov, JETP Lett. {\bf 59},
320 (1994)
\bibitem{Barabanov} A.~L. Barabanov, JETP Lett. {\bf 61}, 7
(1995).
\bibitem{Kievsky} A. Kievsky, S. Rosati, M. Viviani, C.~R.
Brune, H.~J. Karwowski, E.~J. Ludwig, and M.~H. Wood, Phys.
Lett. B {\bf 406}, 292 (1997).
\bibitem{Benenson} W. Benenson, Nucl. Phys. {\bf A588}, 11c
(1995).
\bibitem{Thompson} I.~J. Thompson and M.~V. Zhukov, Phys.
Rev. C {\bf 49}, 1904 (1994).
\bibitem{Glockle} W. Gl\"ockle, {\it The Quantum Mechanical
Few-Body Problem} (Springer-Verlag Berlin Heidelberg, 1983)
pp.\ 27-33.
\bibitem{EH} H. Eikemeier and H.~H. Hackenbroich, Nucl.
Phys. {\bf A169}, 407 (1971).
\bibitem{Kok} L.~P. Kok, Phys. Rev. Lett. {\bf 45}, 427
(1980).
\bibitem{Kamimuratrit} M. Kamimura, Phys. Rev. A {\bf 38},
621 (1988); H. Kameyama, M. Kamimura, and Y. Fukushima,
Phys. Rev. C {\bf 40}, 974 (1989).
\bibitem{Nogga} A. Nogga, D. H\"uber, H. Kamada, and W.
Gl\"ockle, Phys. Lett. B{\bf 409}, 19 (1997).
\bibitem{Kamimura} M. Kamimura, Prog. Theor. Phys. Suppl.
{\bf 62}, 236 (1977).
\bibitem{Merkurev} S.~P. Merkur'ev, Yad. Fiz. {\bf 19}, 447
(1974) [Sov. J. Nucl. Phys. {\bf 19}, 222 (1974)].
\bibitem{cont} A. Cs\'ot\'o, R.~G. Lovas, and A.~T. Kruppa,
Phys. Rev. Lett. {\bf 70}, 1389 (1993); A. Cs\'ot\'o and
G.~M. Hale, Phys. Rev. C {\bf 55}, 536 (1997).
\bibitem{Hale1} G.~M. Hale, R.~E. Brown, and N. Jarmie,
Phys. Rev. Lett. {\bf 59}, 763 (1987).
\bibitem{Hale2} G.~M. Hale and A. Cs\'ot\'o, to be published.
\bibitem{Friar} C.~R. Chen, G.~L. Payne, J.~L. Friar, and
B.~F. Gibson, Phys. Rev. C {\bf 39}, 1261 (1989).
\bibitem{MT} R.~A. Malfliet and J.~A. Tjon, Nucl. Phys.
{\bf A127}, 161 (1969).
\bibitem{Black} T.~C. Black, H.~J. Karwowski, E.~J. Ludwig,
A. Kievsky, S. Rosati, and M. Viviani, Nucl. Phys. {\bf A631}, 680c 
(1998).

\end{references}
\end{document}